\documentclass[journal,comsoc]{IEEEtran}
\usepackage[square,sort,comma,numbers]{natbib}
\usepackage{amsfonts}
\usepackage{acronym}
\usepackage{amsmath}
\usepackage{caption}
\usepackage{subcaption}
\usepackage{float}
\usepackage{comment}
\usepackage{soul}
\usepackage{color}
\usepackage{mathtools}
\usepackage{lipsum}
\usepackage{graphicx}
\usepackage{natbib}
\setcitestyle{square}
\usepackage{graphicx}
\usepackage{balance}
\usepackage{acronym}
\acrodef{vlc}[VLC]{visible light communications}
\acrodef{lifi}[LiFi]{light fidelity}
\acrodef{noma}[NOMA]{Non-Orthogonal Multiple Access}
\acrodef{ris}[RIS]{reconfigurable intelligent surface}
\acrodef{ict}[ICT]{information and communication technologies}
\acrodef{iot}[IoT]{Internet-of-Things}
\acrodef{ioe}[IoE]{Internet-of-Everything}
\acrodef{kpi}[KPI]{key performance indicator}
\acrodef{ai}[AI]{artificial intelligence}
\acrodef{ml}[ML]{machine learning}
\acrodef{los}[LoS]{line-of-sight}
\acrodef{mimo}[MIMO]{multiple-input multiple-output}
\acrodef{wp}[WP]{work package}
\acrodef{em}[EM]{electromagnetic}
\acrodef{adr}[ADR]{angular diversity receiver}
\acrodef{rf}[RF]{radio frequency}
\acrodef{led}[LED]{light-emitting diode}
\acrodef{pd}[PD]{photo-detector}
\acrodef{fov}[FOV]{field-of-view}
\acrodef{ap}[AP]{access point}
\acrodef{pls}[PLS]{Physical Layer Security}
\acrodef{wdm}[WDM]{Wavelength Division Multiplexing}
\acrodef{sir}[SIR]{signal-to-interference ratio}
\acrodef{snr}[SNR]{signal-to-noise ratio}
\acrodef{rgb}[RGB]{red,  green,  and blue}
\acrodef{adr}[ADR]{angular diversity receiver}
\acrodef{adt}[ADT]{angular diversity transmitter}
\acrodef{fpga}[FPGA]{field-programmable gate array}
\acrodef{qos}[QoS]{quality-of-service}
\acrodef{sds}[SDS]{software-defined surfaces}
\acrodef{sm}[SM]{spatial modulation}
\acrodef{csi}[CSI]{channel state information}
\acrodef{rss}[RSS]{received signal strength}
\acrodef{toa}[TOA]{time of arrival}
\acrodef{poa}[POA]{phase of arrival}
\acrodef{aoa}[AOA]{angle of arrival}
\acrodef{sc}[SC]{super position}
\acrodef{sic}[SIC]{successive interference cancellation}
\acrodef{eh}[EH]{energy harvesting}
\acrodef{id}[ID]{information decoding}
\acrodef{slipt}[SLIPT]{ Simultaneous Light-Wave Information and Power Transfer }
\acrodef{sf}[SF]{splitting factor}

\begin{document}

\title{LiFi Through Reconfigurable Intelligent   Surfaces:\\ A New Frontier for 6G? }

\author{~Hanaa~Abumarshoud,~Lina~Mohjazi,~Octavia~A.~Dobre,~Marco~Di~Renzo, \\~Muhammad~Ali~Imran, and Harald~Haas

\thanks{H. Abumarshoud and H. Haas are with the LiFi R$\&$D Centre, University of Strathclyde, Glasgow,  G1 1RD, UK.  (e-mail:\{hanaa.abumarshoud, h.haas\}@strath.ac.uk).}

\thanks{L. Mohjazi   and  M. A. Imran are with the School of Engineering, University of Glasgow, Glasgow, G12 8QQ, UK. (e-mail:l.mohjazi@ieee.org, muhammad.imran@glasgow.ac.uk).}

\thanks{O. A. Dobre is with the Faculty of Engineering and Applied Science, Memorial University, Canada. (e-mail:odobre@mun.ca).}

\thanks{M. Di Renzo is with Universit\'e Paris-Saclay, CNRS, CentraleSup\'elec, Laboratoire des Signaux et Syst\`emes, 3 Rue Joliot-Curie, 91192 Gif-sur-Yvette, France. (e-mail:marco.di-renzo@universite-paris-saclay.fr).}

 }

\maketitle

\begin{abstract}
Light fidelity (LiFi), which is based on visible light communications (VLC), is celebrated as a cutting-edge technological paradigm that is envisioned to be an indispensable part of 6G systems.
Nonetheless, LiFi performance is subject to efficiently overcoming the line-of-sight blockage, whose adverse effect on wireless reception reliability becomes even more pronounced in highly dynamic environments, such as vehicular application scenarios. Meanwhile, reconfigurable intelligent  surfaces (RIS) emerged recently as a revolutionary concept that transfers the physical propagation environment into a fully controllable and customisable space in a low-cost low-power fashion. We anticipate that the integration of RIS in
LiFi-enabled networks will not only support blockage mitigation but will also provision complex interactions among network entities, and is hence manifested as a promising platform that enables a plethora of technological trends and new applications. In this article, for the first time in the open literature, we
set the scene for a holistic overview of RIS-assisted LiFi systems. Specifically, we explore the underlying RIS architecture from the perspective of physics and present a forward-looking vision that outlines potential operational elements supported by RIS-enabled transceivers and RIS-enabled environments. Finally, we highlight major associated challenges and offer a look ahead toward promising future directions.

\end{abstract}

\section{Towards Intelligent  LiFi}
Harnessing the power of billions of connected devices, future \ac{ioe} is envisioned to enable innovative and progressive  services  such as  telemedicine, extended reality and  automatic  high-precision  manufacturing.  Such    applications  require unprecedented wireless connectivity with specific \acp{kpi} including  terabit-per-second speed, exceedingly high-reliability,  extremely low-latency and high energy-efficiency, which cannot be efficiently supported by  $5$G networks.  To cater for  these demands, $6$G is envisioned to be  built  on a new physical layer architecture that incorporates subterahertz and visible light bands to support and complement \ac{rf} communications.

\Ac{lifi}, which offers a fully networked bidirectional wireless  solution  based on \ac{vlc},  has been identified as an important component in the $6$G  blueprint,  with predictions of the \ac{lifi} market reaching  $\$8$ billion by $2030$ \cite{market}.   The  directionality  and   short-range travel distances of  light signals  allow for extreme cell densification in indoor, vehicular and underwater communications, providing  secure high-speed connectivity.  It is also considered an energy and cost effective technology as it utilises  the lighting infrastructure to provide connectivity on top of the illumination functionality.  Nonetheless,  the achievable capacity in \ac{lifi} systems is limited by: 1) the  modulation bandwidth of the transmitting \acp{led}, and  2) the high dependency on the \ac{los}  which means that signal quality is influenced  by link blockage and random receiver orientation.  The use of \ac{mimo} configurations is a promising solution to  overcome these limitations and boost the capacity and diversity gains of \ac{lifi}, albeit with a possibly hindered performance  due to the  high correlation between  the spatial subchannels.

Until recently, the design  of  \ac{lifi} systems  was  mainly focused on enhancing  the  transmission and reception capabilities  in the face of  undesirable uncontrollable   channel conditions. In order  to achieve the \acp{kpi} of $6$G, there is a need to go beyond the current communication architecture, which is limited by the transceiver front-ends capabilities, and to exploit the available degrees of freedom in  the environment. Based on this, intelligent \ac{lifi} can autonomously and dynamically adapt its operation in order to achieve ubiquitous connectivity that is resilient to   adverse probabilistic effects such as blockage and random receiver orientation.  This article aims to give a forward-looking vision of the role that \acp{ris} can play in enabling  intelligent   \ac{lifi} as a  new frontier for $6$G  wireless networks. To the best of our knowledge, this is the first article that attempts to give an overview of the opportunities, applications, and challenges related to the integration of \acp{ris} in the context of \ac{lifi}.

\subsection{RIS: A Disruptive  Concept}
A \ac{ris} comprises a  metasurface that can be proactively reconfigured to  alter the wireless wave propagation. The \ac{em} response of each \ac{ris} element can be adjusted by tuning the surface impedance through electrical voltage  stimulation. 
Based on this newly emerging  concept,  future wireless networks may utilise the different  surfaces and physical objects in the environment   as wireless boosters to enhance their operations.
\ac{ris} research   has recently  become  a focal point of interest  in wireless communications communities as it offers a spectrum, energy, and cost efficient approach for  sustainable evolution in wireless systems.    The  integration of \ac{ris} in \ac{rf} communications  is shown to offer  performance enhancement in terms of coverage, \ac{qos}, and security \cite{9140329}.
The application of this emerging technology in optical communications is  still in its infancy. The use of \ac{ris} in free space optical systems was investigated as a means to maintain the \ac{los}  in the existence of  link obstructions in \cite{9013840},
while  \cite{9276478} and   \cite{9348585} considered the use of  wall-mounted \acp{ris}  to focus the incident light beams towards  optical receivers in  indoor \ac{vlc} systems.

To better understand the potentials and limitations of the interplay between \acp{ris} and \ac{lifi}, we next shed light on some of the reported capabilities of metasurfaces to control visible light propagation.

\subsection{Metasurfaces From the View of Physics}
\begin{figure}[h]
	\centering
	\resizebox{1\linewidth}{!}{\includegraphics{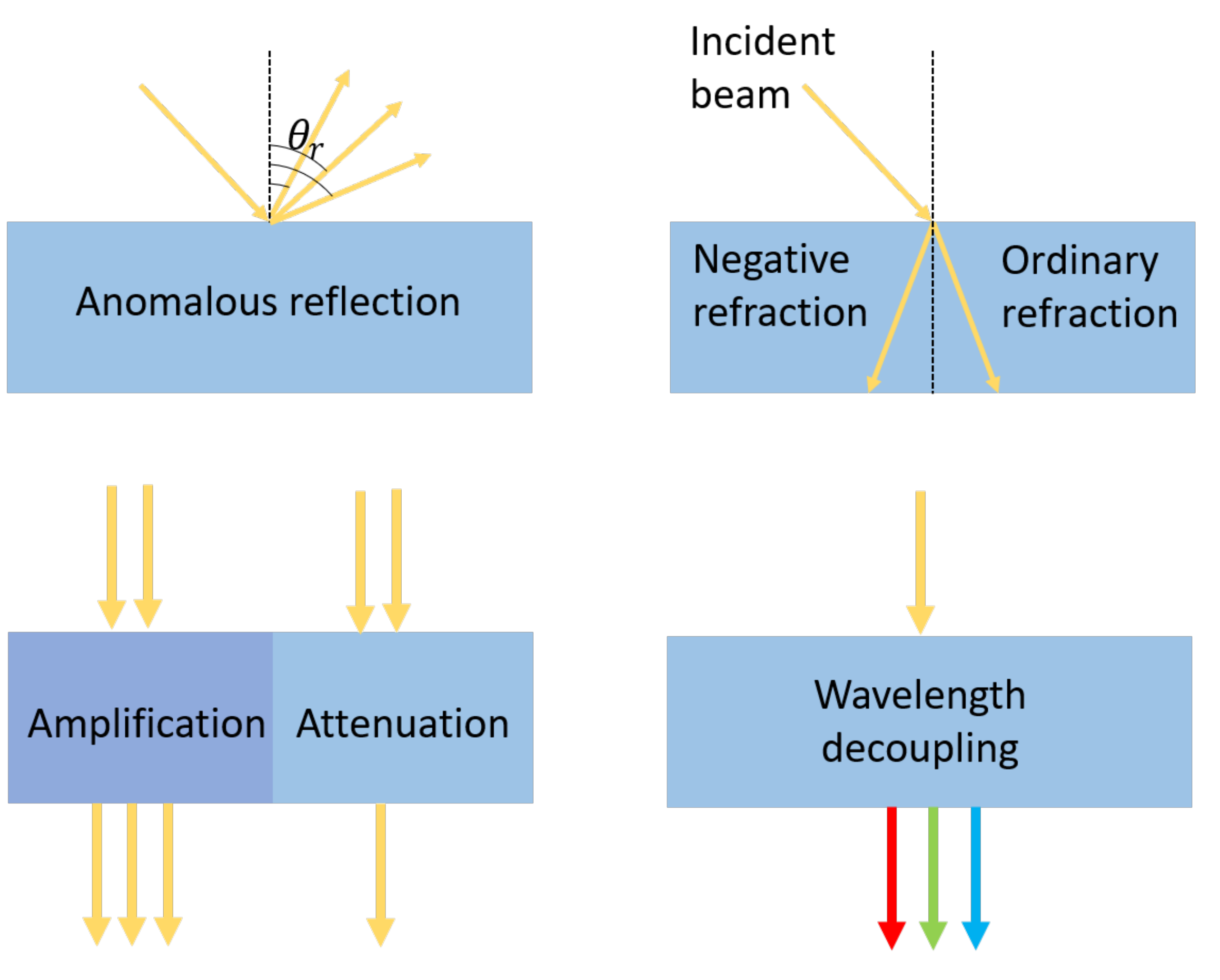}}
	\caption{ Metasurfaces light manipulation capabilities. }
	\label{fig:functionalities}
\end{figure}

Metasurfaces  are two-dimensional artificial structures that consist of programmable   sub-wavelength metallic or dielectric elements whose  \ac{em} characteristics  can be reconfigured  to  introduce an  engineered response to the  incident wave-front by  manipulating   the outgoing photons.
Although the use of metasurfaces in optical communications have only recently became an active area of research, their capabilities in  light manipulation have already been  developed in the field of flat optics. Fig. \ref{fig:functionalities} illustrates  four of the most important light manipulation capabilities that can be realised by metasurfaces and are detailed as follows:

\begin{enumerate}
    \item Refractive index tuning: with a negative  refractive index it is possible to reverse the  phase velocity of the light-wave and bend it in a direction that is impossible with
a positive index \cite{refractive}. 
\item Anomalous reflection:
metasurfaces make it possible to break the law of reflection and
steer the light  into desired directions with near-unity efficiency  \cite{reflection}.
\item Signal amplification and attenuation:
metasurfaces can be tuned  to provide light amplification, attenuation, or even complete absorbance \cite{9354893}.
\item Wavelength decoupling:
it is possible to engineer wavelength-specific \ac{em} response  to  control different wavelengths  independently  \cite{OAM}.
\end{enumerate}

Based on these capabilities, we next  draw a vision on the use of metasurfaces within  \ac{lifi} transceiver structures.

 \begin{figure*}
	\centering
	\resizebox{0.9\linewidth}{!}{\includegraphics{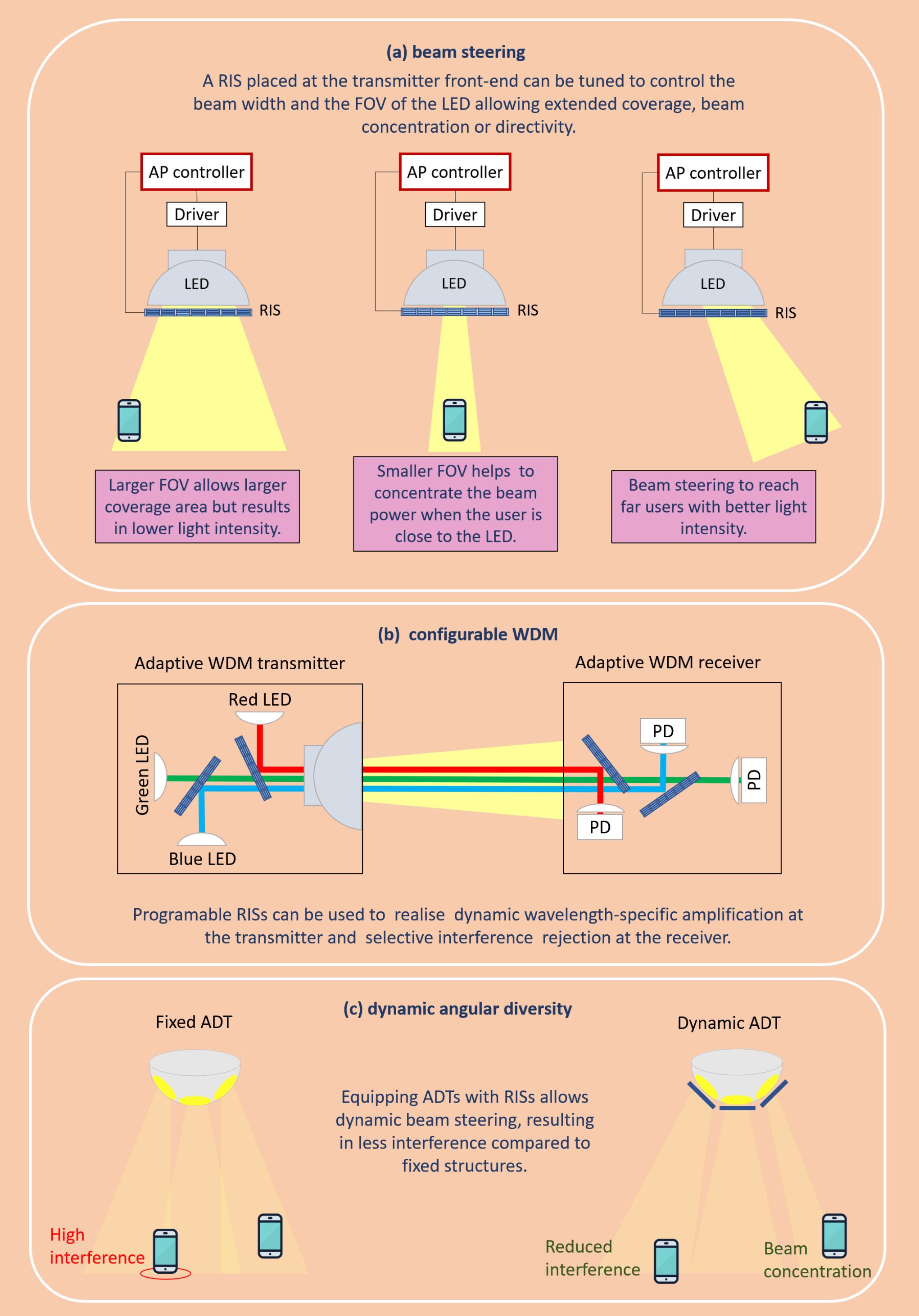}}
	  \caption{Envisioned functionalities for \ac{ris}-enabled transceiver front-ends. }
    \label{fig:TxRx}
\end{figure*}

\section{RIS for Adaptive Transceivers}
\Ac{lifi} transmitters  emit directional beams of light characterised by  their  \acp{fov}.
A large \ac{fov} results in a  wider beam angle, which  is beneficial in terms of providing higher  coverage and  more uniform illumination. This comes at the cost of producing lower light intensity at the receiver, i.e. lower Watt/unit area. As a result,   transmitters with small \acp{fov}  can  better preserve the received optical power,  albeit requiring perfect link alignment.

Light rays falling within the receiver's  \ac{fov} generate an electrical current that is directly proportional to the  received optical power. The more data-carrying photons are observed at the \ac{pd} surface, the higher the  detection accuracy  is.  This suggests that using \acp{pd} with large physical areas  leads to  better  performance. However, large \acp{pd}  typically have lower  $3$ dB bandwidth due to the increased capacitance, and hence, are not suitable for high data-rate applications. To tackle this problem,  receivers employ \acp{pd} with small physical areas combined with convex, spherical,  or compound parabolic concentrating lenses, to help  focus the light rays that fall in the receiver's \ac{fov}  onto the \ac{pd} surface.

The  optical components used in  state-of-the-art \ac{lifi} transceivers  are   non-configurable structures, meaning that their   characteristics,   such as  \acp{fov}, amplification factors and  operating wavelengths,  are predetermined and   cannot be dynamically adjusted. Due to the tunability of their physico-chemical characteristics,  \acp{ris} can  unleash the possibility to achieve  dynamically tunable transceivers as  discussed in the following.

\begin{figure*}
	\centering
	\resizebox{0.9\linewidth}{!}{\includegraphics{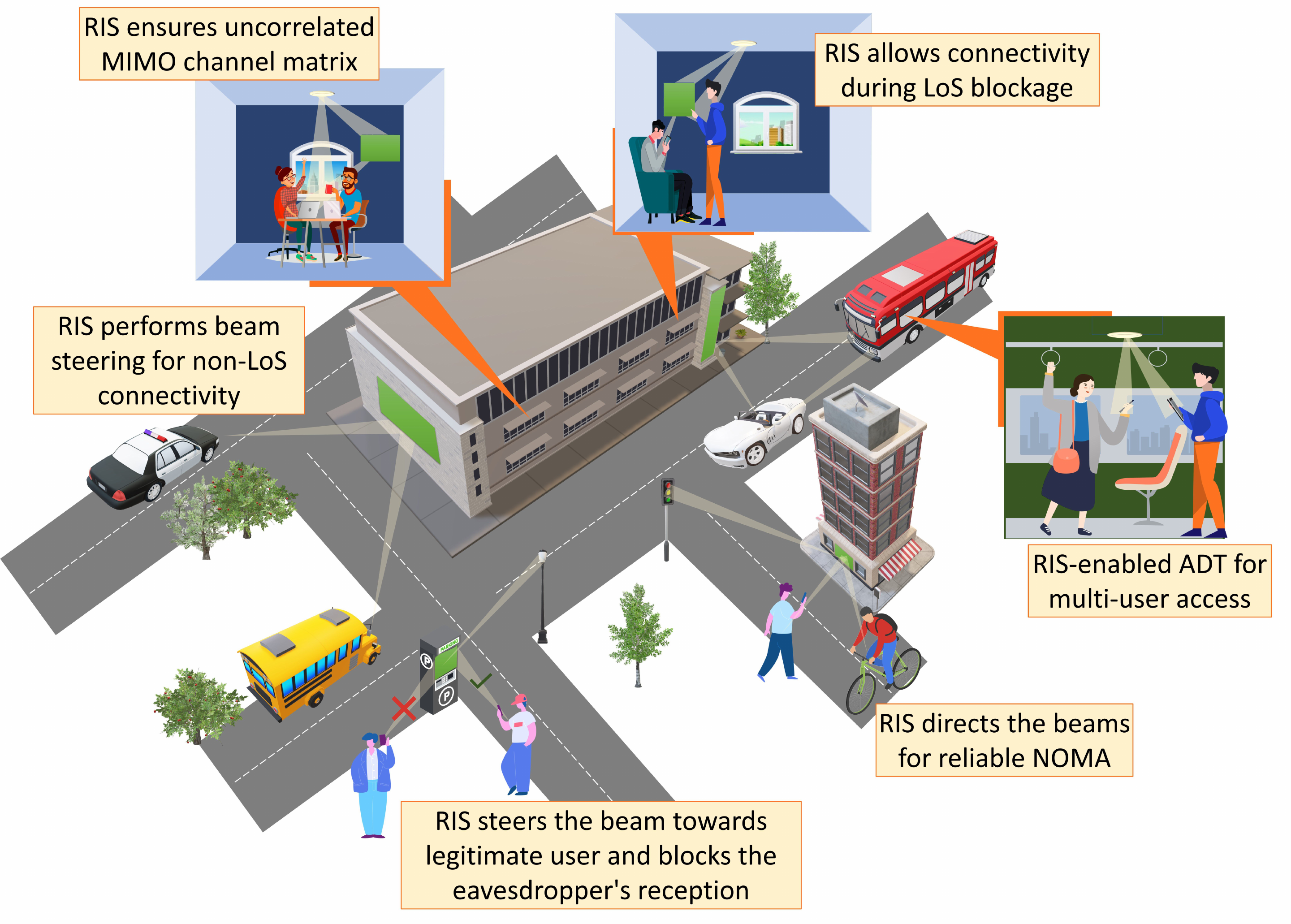}}
	  \caption{Envisioned \ac{ris}-enabled smart \ac{lifi} applications in indoor and outdoor scenarios.  }
     \label{fig:App}
\end{figure*}


\subsection{Beam Steering and Amplification}\label{beam}
A concept design of a \ac{ris}-based  receiver was demonstrated in \cite{9354893},  where the traditional convex lens is  replaced with a  liquid crystal array to  provide a two-fold impact on the incident beam, namely: steering and amplification, which was shown to extend the coverage range by few meters.
The amplification in this case is achieved locally using the surface waves, but the \ac{ris}  is  globally passive, i.e. it does not require power supply.
In such configurations, it is critical that the \ac{ris}  handles  amplification without driving the  \ac{pd} to work  in the saturation region  to avoid signal  distortion. Besides enabling beam steering at the receiver, it is also possible to exploit this functionality  at the transmitter to enhance the beams' directivity, as illustrated in  Fig. \ref{fig:TxRx} (a).

\subsection{Configurable \ac{wdm} Transceivers}
In \ac{wdm}, data streams are multiplexed on distinct wavelengths by utilising a  combination of different  coloured \acp{led}.
Dichroic mirrors  are typically used  in \ac{wdm} transceivers  to combine  the different wavelengths at the transmitter and to decorrelate the received signals at the receiver.
Such transceiver  structures are fixed configurations that can only work for a  specific \ac{wdm} scheme, i.e. a receiver comprising \ac{rgb} filters can only decorrelate   an \ac{rgb}-based \ac{wdm} signal. Since the \ac{em} response of \acp{ris} can  be tuned at  different wavelengths, they could be possibly integrated in \ac{wdm} transceivers  as illustrated in Fig. \ref{fig:TxRx} (b).   The  spectral width  of \ac{ris} elements can then be tuned to realise  dynamic  interference filtering according to the used transmission scheme.   \ac{ris}  tuning allows for allocating different weights to each received wavelength, and hence,  the effective  signal-to-interference ratio  can be optimised.
The realisation of \ac{wdm} tunable transceivers  requires the fabrication and characterisation of metasurfaces whose response can be finely tuned for different  wavelengths.

\subsection{Dynamic  Angular Diversity}
Fig. \ref{fig:TxRx} (c) illustrates  an \ac{adt} consisting of multiple \acp{led}, each pointed towards a particular direction to create  multiple   sub-cells for  multi-user access.  Since the optical components in \acp{adt} are closely located, the degree of decorrelation between the   subchannels is not always high enough to provide  meaningful diversity gain. Interference levels can be high for users existing in the overlapping area of two sub-cells, which diminishes the performance gain. There is a need to adapt the spatial separation and width of the optical beams according to the  number and locations of users, since fixed transceiver structures are not always guaranteed to provide the required performance gain. To tackle this issue,  \ac{ris} may be utilised to dynamically change the width and divergence of the optical beams, allowing  higher numbers of users to be  simultaneously served with less interference and  higher area spectral efficiency.

\section{ \ac{ris}-enabled Environments}
In this section, we discuss some potential applications for the integration of \ac{ris} within  \ac{lifi} environments, as  illustrated in Fig. \ref{fig:App}.

\subsection{Non-\ac{los} Links for Resilient Connectivity}
\begin{figure}[h]
	\centering
	\resizebox{1\linewidth}{!}{\includegraphics{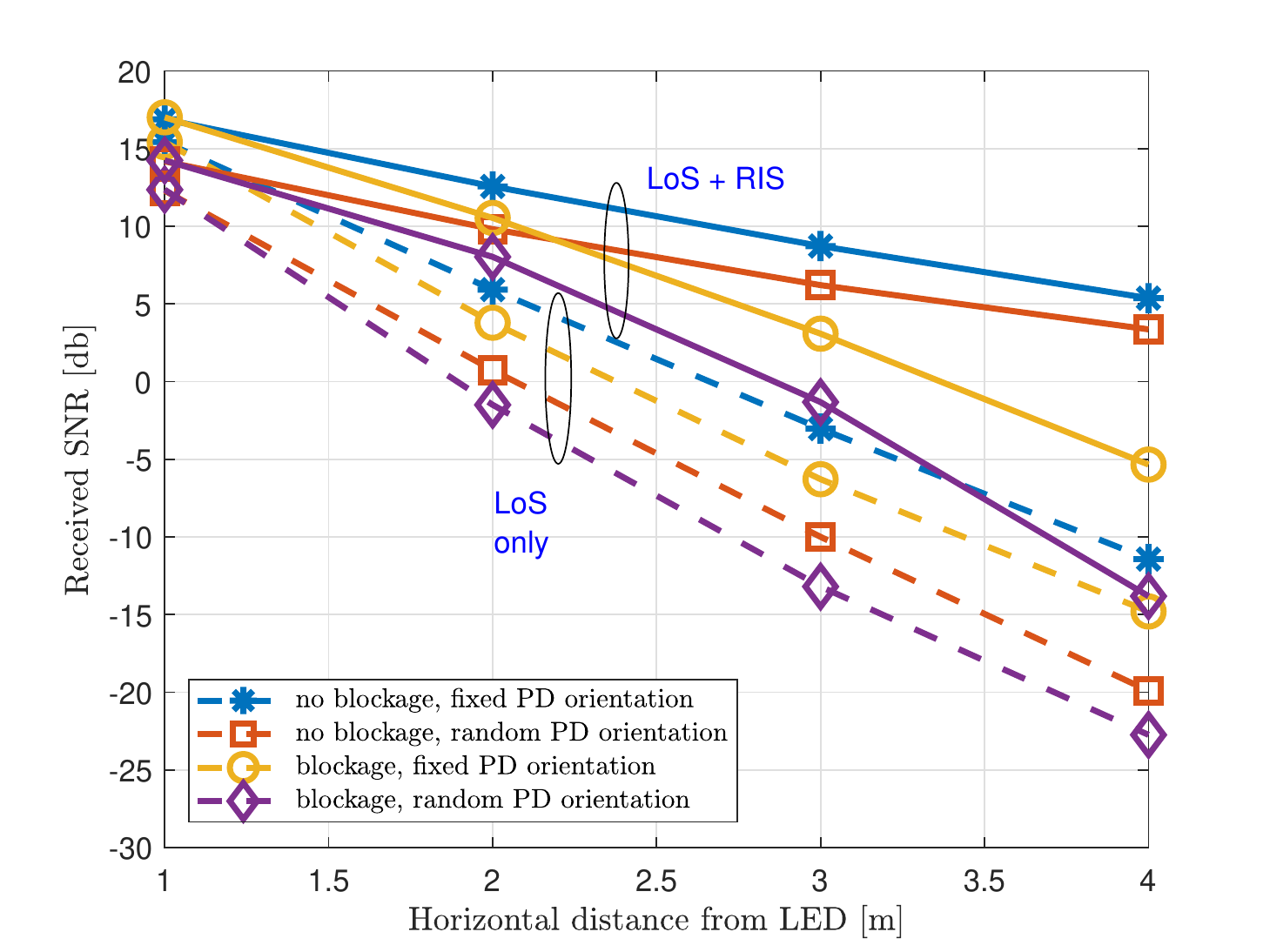}}
	\caption{Simulation results for received signal-to-noise ratio (SNR) vs. the horizontal separation between the user and the LED in  a $8\times8\times3$ m$^3$ room setup with $1$ m$^2$ \ac{ris} mounted on one wall.
}
	\label{fig:blockage}
\end{figure}

The optical  link quality   primarily relies on the existence of \ac{los} paths and is  likely to be significantly deteriorated or completely interrupted in the event of \ac{los} obstruction, which necessitates  perfect alignment between the transmitter and the receiver.  While this is a viable solution in \ac{vlc} links between static terminals, the case is more complicated in \ac{lifi} networks supporting high user mobility.
\ac{lifi} receivers  perceive multi-path components that are coherently added at the \ac{pd} to maximise the total received power.   In typical indoor environments, however, non-\ac{los} signals are totally uncontrolled and nearly isotropic in their spatial distribution and are very weak in their average intensity. 
It was shown in \cite{9276478} that using  a $25$ cm × $15$ cm reflector can enhance the received power five fold compared to the direct \ac{los} link, which makes  \ac{ris} a very appealing solution to compensate for \ac{los} blockage in \ac{lifi} systems. 
Leveraging \ac{ris} could indeed  enable  ubiquitous connectivity    under  mobility conditions and link blockages as well as random device orientations, as shown in Fig. \ref{fig:blockage}, making \ac{lifi} resilient to these probabilistic factors.

\subsection{Uniform Coverage for Seamless User Experience }
\begin{figure*}
     \begin{subfigure}[b]{0.5\linewidth}
         \includegraphics[width=\textwidth]{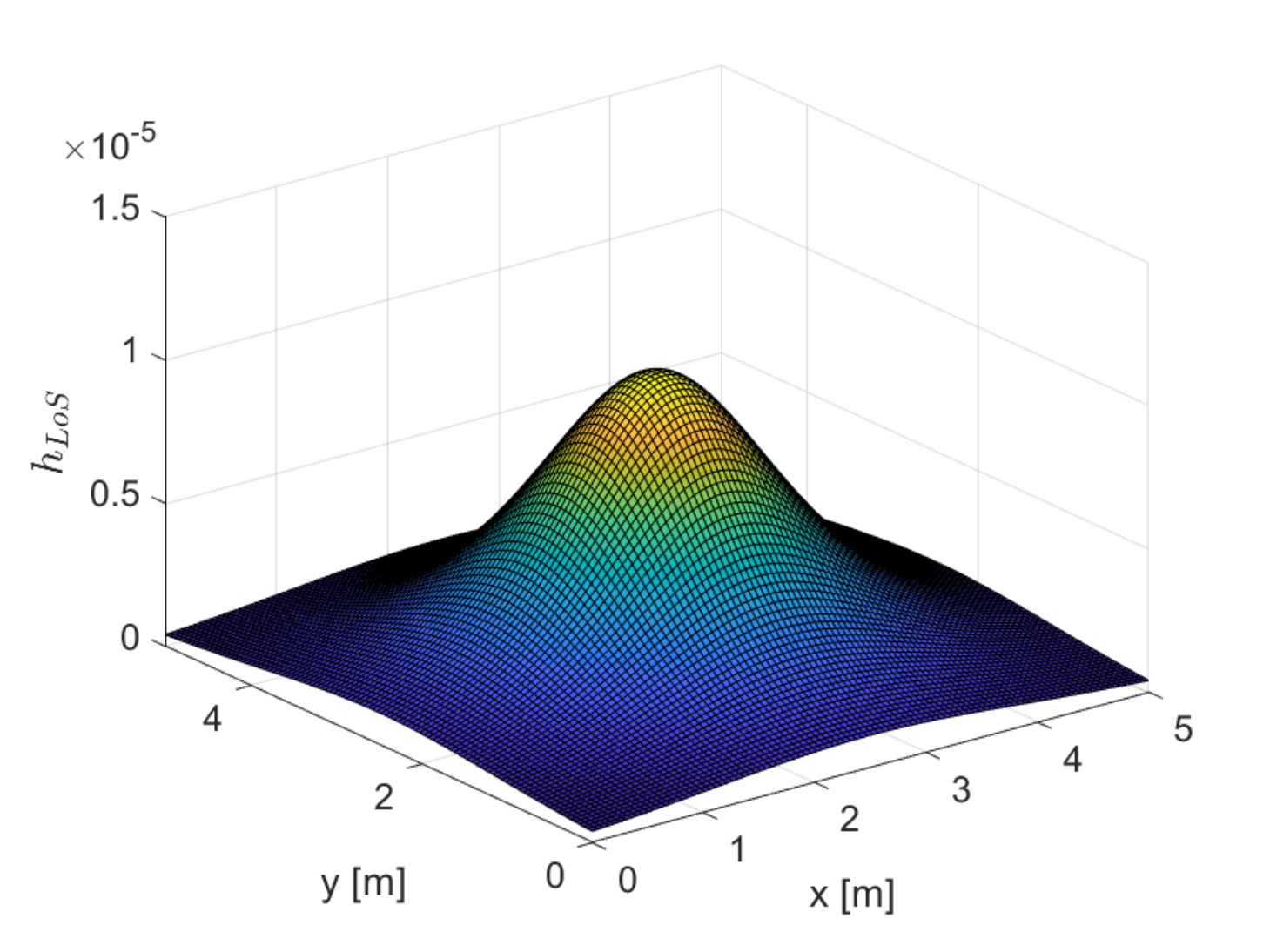}
         \caption{LoS only.}
          \vspace{0.1 cm}
         \label{fig:h_los}
     \end{subfigure}
     \begin{subfigure}[b]{0.5\linewidth}
         \includegraphics[width=\textwidth]{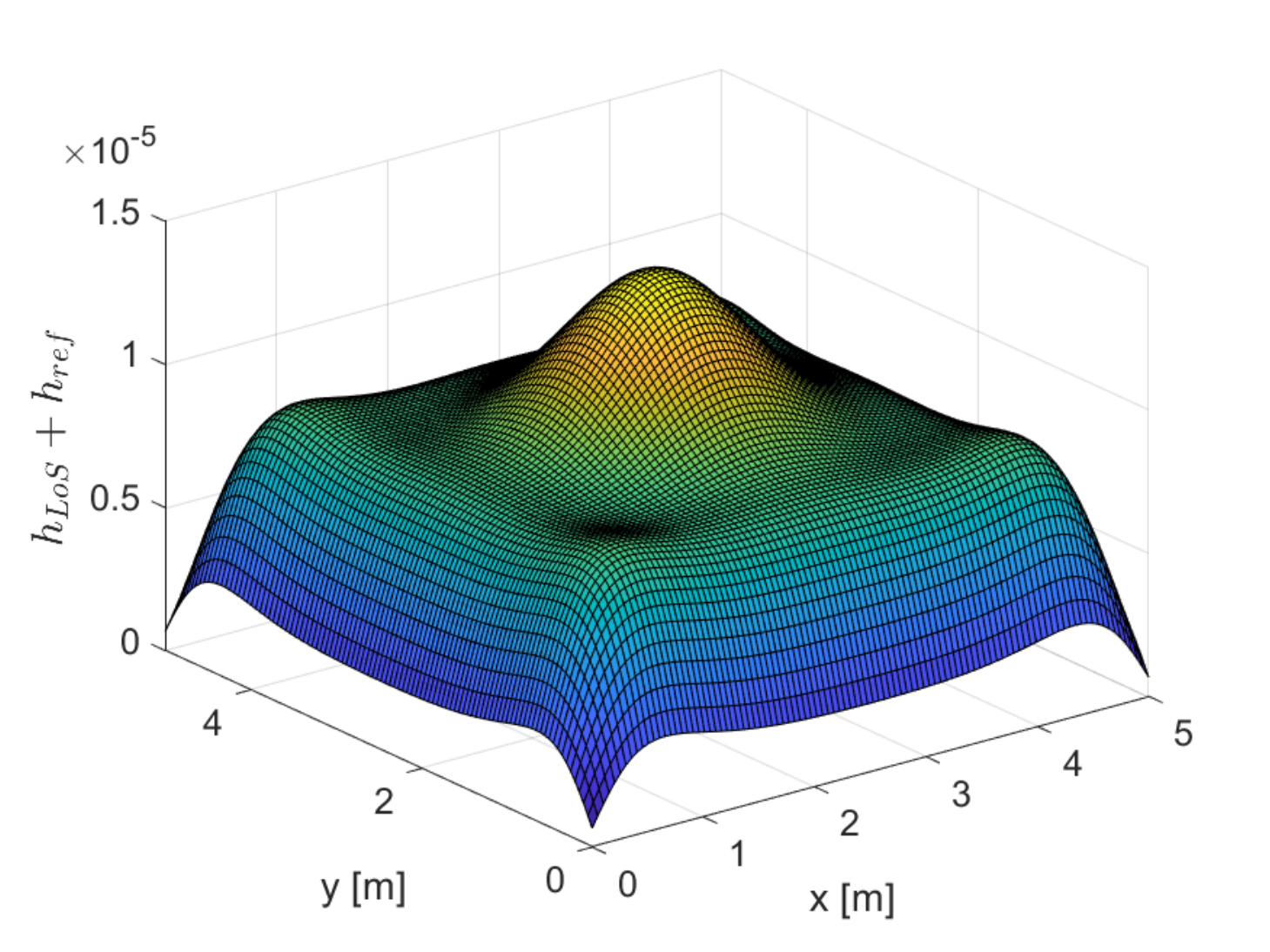}
         \caption{LoS + RIS reflected signals.}
         \vspace{0.1 cm}
         \label{fig:h_total}
     \end{subfigure}
     \caption{Simulated channel gain distribution across a $5\times5\times3$ m$^3$ room with a) LED LoS only, and   b)  LED assisted with \ac{ris}  covering an area of $1$ m$^2$ on each of the four walls. }
     \label{fig:channel}
\end{figure*}

\ac{lifi} coverage  is typically confined within the width beam of the transmitted light which spans  $2$ m - $3$ m in diameter.
While this property  makes \ac{lifi} particularly secure against eavesdropping, it  results in short  communication range and non-uniform distribution across the area, as shown in Fig. \ref{fig:channel} (a). The reduced link reliability at the cell-edge means that users moving in the proximity of the \ac{lifi} \acp{ap} will need to perform frequent horizontal handovers, between neighbouring \ac{lifi} \acp{ap}, or vertical handovers, between the \ac{lifi} and the \ac{rf} networks.
\ac{ris}-enabled   environments can enhance the \ac{lifi}  coverage by  mitigating  dead-zone problem for cell-edge users as shown in  Fig. \ref{fig:channel} (b). For a  user  located at the room edge, the direct  received power from an  \ac{ap} mounted in the room centre could range from low to none. However, since the user is close to the wall-mounted \ac{ris}, the reflected signal strength can allow for meaningful link quality.

\subsection{Spatial Decorrelation for Enhanced MIMO}
The application of \ac{mimo} systems in \ac{lifi} is particularly  appealing,  due to the existence of large numbers of \acp{led} in  a single luminary and the existence of many luminaries in various indoor and outdoor settings.  Unlike \ac{rf},  optical  links do not exhibit fading characteristics and thus, do not always represent unique signatures. In fact, the optical channel is subject to a strong deterministic component, since it is  mainly influenced  by the transceivers'  geometry, meaning that  the \ac{mimo} channel can be rank-deficient or  ill-conditioned. This constitutes a performance  bottleneck for different \ac{mimo} configurations such as spatial  multiplexing and spatial modulation,  particularly  in  massive \ac{mimo}  due to the high proximity of the  front-ends.

One  possible solution to tackle this problem is  to   carefully align
the \acp{led} and \acp{pd} in a way that ensures uncorrelated subchannels.  However, optimising the locations of the transceiver front-ends in \ac{lifi} systems that support high user mobility proves to be a challenging requirement.
\ac{ris}-enabled  environments could offer an excellent solution to improve the \ac{mimo} channel matrix rank,   yielding better spatial decorrelation. In scenarios where \ac{mimo} receivers observe equivalent subchannels from the direct \ac{los} link,  the intensity of the reflected paths from \ac{ris} to receivers can be controlled so that the total observed subchannels are sufficiently distinguishable. Better  spatial uniqueness will enable  higher spectral efficiency and boost the achievable data rates of  \ac{lifi} \ac{mimo} systems.

\subsection{\Ac{ris}-Based Modulation}
\begin{figure}[h]
	\centering
	\resizebox{0.9\linewidth}{!}{\includegraphics{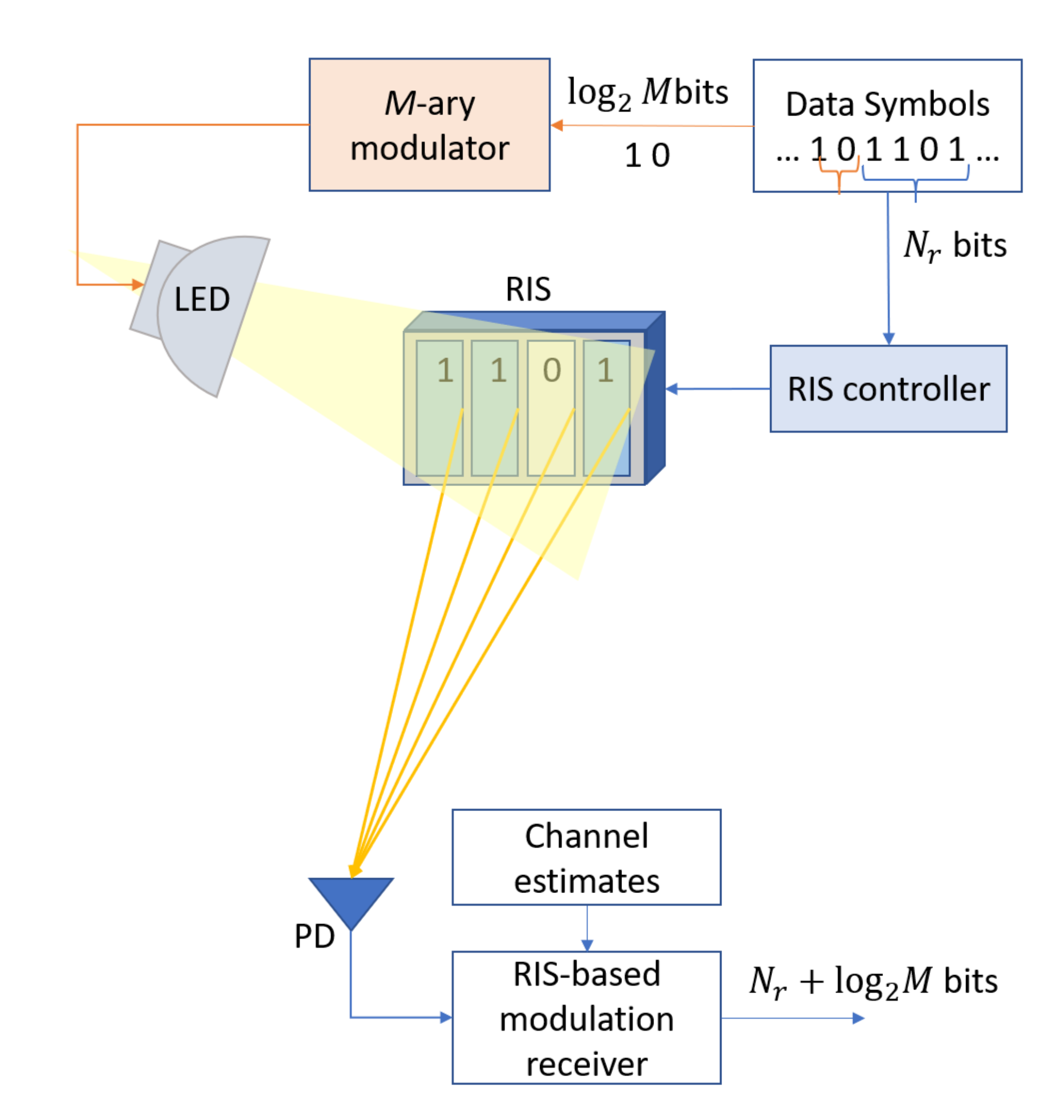}}
	\caption{\ac{ris}-based modulation in a  \ac{lifi} transceiver equipped with  RIS containing $N_r$ reflecting elements. }
	\label{fig:mbm}
\end{figure}

A \ac{ris}   can be used to provide  an additional dimension for data modulation by controlling the \ac{em} response of each of its reflecting elements \cite{ris-mod}.  \ac{ris}-based modulation is a generalisation of \ac{sm} that exploits the radiation patterns of reconfigurable elements instead of the activation states of \acp{led}. Compared to  conventional \ac{sm}, which utilises multiple \acp{led},   \ac{ris}-based modulation  offers a smaller size, lower cost, and  more flexible solution to increase the system spectral efficiency.
The operating principle of  \ac{ris}-based modulation  is illustrated in Fig. \ref{fig:mbm}.  A \ac{ris} equipped with multiple  reflecting elements receives   the data modulated signal from the \ac{led}  and reflects it to the intended receiver.  Assuming that each reflecting element has two distinct reflecting states (on/off), different on/off combinations correspond to different observed channel realisations at the receiver. Hence, by controlling the \ac{em} response of $N_r$ reflecting elements,  a spectral efficiency enhancement of $N_r$ bits per symbol is realised.  Assuming that the \ac{led} transmitted signal is $M$-ary modulated, the total spectral efficiency of the system is $N_r+\log_2(M)$.

In principle,  the spectral efficiency of  \ac{ris}-based modulation  increases linearly with the number of reflecting elements. However, the  error performance   is practically limited by  the Euclidean distance between the   constellation points, which is governed by the degree of decorrelation among the reflections. In other words, although a high number of distinct channel combinations  can be generated, a superior performance enhancement  can be only realised  if these combinations are clearly distinguishable.  At the receiver terminal, the detector task is not only to decode the $M$-ary modulated symbol, but also to estimate the channel state of each reflecting element. This implies that the receiver needs to be trained with pilot signals from all possible \ac{ris} combinations. While this task might be complex in  fading \ac{rf} channels, the complexity of obtaining the \ac{csi} is considerably lower in \ac{lifi} due to the somewhat deterministic nature of the optical channel.
An important question that arises here is: how will the \ac{ris} receive the intended data bits in order to alter the reflecting elements' states? A possible way to do this is to mount the \ac{ris}  on a fixed object that is connected to the \ac{lifi} backhaul network.  For example, an indoor  wall-mounted \ac{ris}  can be connected to the same power-line communication link feeding  the  \acp{led} in the ceiling. The same concept can be applied  for an outdoor scenario in which  \ac{ris} is   mounted on a traffic light to modulate the light signals and reflect them to vehicular  receivers.
\subsection{\Ac{noma}}
Power-domain \ac{noma} is a spectral  efficient  technique in which  different users are simultaneously  served using a single resource block, i.e., accessing the full time and frequency resources.
The spectral efficiency  gain offered by \ac{noma} is  dependant on  having
distinct channel  conditions for different users, which is not always guaranteed in \ac{lifi}. In fact, since the optical channel gain is determined almost exclusively by the relative location of the user with respect to the \ac{ap}, it is highly likely that multiple users will have similar channel conditions, hindering the feasibility of  \ac{noma}.

The case might completely change in \ac{ris}-assisted  \ac{lifi} systems. By controlling the multi-path  channel propagation, it would be possible to dynamically  alter the perceived channel gains at different users to achieve  performance enhancements  in two ways:
\begin{enumerate}
    \item Higher reliability: by introducing and controlling the channel gains, it is possible to create perfect conditions for effective power allocation, and thus,  successful  successive interference cancellation.
    \item Enhanced  fairness:  \ac{noma} users with lower decoding order, i.e., lower channel gain, have to always decode their signals with the existence of interference which means that their achievable data rates might not be enough to satisfy  their \ac{qos} requirements. Dynamic \ac{ris} tuning makes it possible to change the users decoding order despite their locations,
    leading to enhanced fairness.
\end{enumerate}

\subsection{\Ac{pls}}
Although \ac{lifi} is  inherently secure in confined places  because light signals do not penetrate walls, \ac{lifi} links are susceptible to eavesdropping by malicious users existing under the coverage area of the same AP. This makes it particularity crucial to secure \ac{lifi} transmissions in  public spaces,  such as shopping malls, libraries,  and airports, as well as in outdoor and vehicular applications  \cite{PLS}.
The integration of \ac{ris} in \ac{lifi} systems can lead to enhanced \ac{pls} in one of the following ways:
\begin{enumerate}
    \item Enhanced secrecy capacity:  dynamic multi-path tuning to maximise  the channel gain for legitimate  users while minimising it for the eavesdroppers.
    \item Jamming:  randomised multi-path reflections directed towards the eavesdropper to produce artificial noise.
    \item Secure beamforming: multiple \ac{ris} elements can be configured to produce precoding vectors  so that the data signal can only be decoded at the legitimate user.
\end{enumerate}
In order to realise \ac{ris}-based \ac{pls}, there is a need to  investigate the secrecy capacity performance in such systems in relation to   the locations and capabilities of \acp{ris}, particularly with the existence of ambient noise in outdoor links,    and to   understand how \acp{ris} can be enabled to perform users authentication.

\section{RIS Beyond Communications}
In the following, we discuss how  \ac{ris}-enabled \ac{lifi} systems have the potential to provide enhanced monitoring and power transfer, on top of their communications functionalities.

\subsection{\ac{ris}-Aided Light-based Monitoring}
The dynamics of light propagation and reflection in \ac{lifi} systems carry huge amounts of information about the distance travelled by the light rays, the nature of obstacles encountered, and the movement patterns in the coverage areas.  
Light-based sensing solutions, such as  LiSense, StarLight, and Li-Tect (\cite{8520863} and the references therein),  proved  that light has the potential to provide high-accuracy 3D  sensing. In these techniques,  propagation paths of beacon signals between multiple optical transmitters and receivers  were  utilised to reconstruct point clouds for the purpose of localisation.


In \ac{ris}-assisted environments, each reflected path  constitutes a unique spatial signature depending on the locations and \ac{em} response of the objects existing in the environment.  The reflections from the \ac{ris} elements can be shaped by controlling the \ac{em} response of each element in order to enable better mapping from the position space  to the  measurement space. Based on this, accurate localisation can be achieved for indoor applications, including   physical rehabilitation, gesture recognition, and automated manufacturing, as well as   for  outdoor applications, such as measuring the distances between vehicles, pedestrians,  and buildings for safety and surveillance.

\subsection{\ac{slipt}}
The potential of \ac{lifi} systems to provide illumination while supporting information transfer concurrently with \ac{eh} has led to the emergence of \ac{slipt} as an innovative approach to extend the lifetime of energy-constrained terminals \cite{Pan}. In this context, visible light-waves broadcasted through \acp{led}  are captured  at the receiver to perform \ac{id} and \ac{eh} simultaneously according to a pre-defined time or power \ac{sf}.  \ac{slipt} relieves the bottleneck of energy-sensitive networks, while avoiding the safety problems imposed by traditional wireless \ac{rf} power transfer systems, and is, thus, appealing in many indoor \ac{ioe} scenarios. One of the key challenges in \ac{slipt} is that transceivers should be capable of identifying an optimal \ac{sf} to satisfy the demands of both \ac{eh} and \ac{id}. Furthermore, the \ac{eh} efficiency is directly impacted by the light collection area at the receiver. The merits of integrating \ac{ris} within \ac{lifi} transceivers can be leveraged to dynamically tune the transceiver characteristics, through beam steering and amplification, as described in Sec.~\ref{beam}, which together with smartly optimising the \ac{sf} can lead to a balance between \ac{id} and  \ac{eh}. The synergy of \ac{mimo} and \ac{vlc} creates a promising platform for highly efficient \ac{slipt}-enabled systems, albeit at the cost of requiring complex signal processing at both source and receiver \cite{Pan}. 
We envision that \ac{ris}-assisted smart environments can potentially offer an on-demand efficient resource allocation for such systems, especially in dense deployments. Capitalising on the design of efficient joint active and passive beamforming techniques, the amplitudes and phases of independently transmitted and reflected signals can be configured such that they are added constructively in desired directions and destructively in the undesired ones to create multiple data/light-wave energy streams conveyed to multiple receivers. This delivers an unprecedented flexibility of multiple access control in conjunction with transmission rate and light transmit power control that maintain individual  \ac{qos} requirements on \ac{id}/\ac{eh} and guarantee inter-user interference mitigation in a dynamic fashion. 

\section{Challenges}
In this section, we discuss some critical challenges that need to be addressed for realising the full potentials of integrating \ac{ris} in \ac{lifi}.

\subsection{Modelling and Characterisation}
The development of realistic and accurate channel models is essential to capture the fundamental performance limits of \ac{ris}-assisted \ac{lifi} systems.
Moreover, there is a need to quantify the efficiency  and response time to  achieve  specific functionalities, such as  amplification, absorption,  anomalous reflection, etc.
It is noted that the use of \ac{ris}-assisted \ac{lifi} is  particularly  desirable in outdoor   transmissions,  such as vehicle-to-vehicle and vehicular-to-infrastructure,    due to the requirement of extremely small beam angles in such long-distance links.   Underwater \ac{lifi} systems could also hugely benefit from dynamic tuning and beam steering  because the  underwater  environment is characterised by severe  signal attenuation, and  thus, any possibility to precisely steer and focus the light beams will undeniably bring advantages. Accordingly, it is critical to understand how \acp{ris} perform  in different media,  such as underwater and outdoor environments with high ambient noise. Comprehensive practical implementations and measurements over practical setups are required to asses  the performance  in real-world scenarios.

\subsection{Real-Time Estimation and Optimisation}
\acp{ris} provide the fabric for dynamic \ac{lifi} configurability based on the knowledge of
the varying system parameters, which necessitates the  availability  of accurate \ac{csi} of direct and non-direct channel paths  of all network users.
The natural questions that arise are:  how will   the necessary amount of data be acquired in a practical and cost-effective manner, how to integrate the optimisation and resource management capabilities into the network topology, and what is the convergence, stability,  and required frequency of this dynamic sensing and reconfiguration.  Data  acquisition and optimisation can be possibly performed in a distributed or centralised manner. In the  distributed approach,  \acp{ris} can locally  estimate the channel gains and autonomously reconfigure their \ac{em} response accordingly,  if they comprise sensing and processing capabilities. The use of  \ac{ris} arrays with high numbers  of elements and complex functionalities will require complex optimisation, resulting in increased computational, energy,  and cost overhead. The  centralised approach, on the other hand, employs a central control unit at the \ac{ap} which executes the estimation and optimisation protocols and then communicates the optimal design coefficients to the \ac{ris}'s control layer. The obvious advantage of the centralised approach is lower energy consumption and simpler hardware design, while the main challenge   is the occurring  signalling overhead  between the \ac{ap} and the \ac{ris}, which would be more pronounced in high mobility scenarios. The trade-off between distributed and centralised approaches needs to be quantified so as to develop practical strategies for different  applications.

\section{The Way Forward}
In the following, we present a forward-looking discussion on promising future directions.
\subsection{Interplay of \ac{ml} and RIS-Assisted LiFi}

The promising prospects of \acp{ris} provide a foundation to support seamless real-time adaptivity of the LiFi-enabled wireless environment. However, their operation is intertwined with the need to implement a complex level of coordination to sustain a desired holistic behaviour while ensuring scalability and overhead reduction. This complexity stems from the fact that turning the environment into a programmable,  and a partially deterministic space,  relies on incorporating a colossal number of parameters in the performance optimisation, which in  turn depends on a large amount of sensed data that is processed in a time-critical manner to interact with and incorporate responses
from the dynamic environment surrounding \acp{ris}. This suggests that network resource management would be an extremely challenging task, since memory, computing power, traffic demands, space, and so forth are to be considered. To tackle this challenge, \ac{ml} solutions may be exploited to support \ac{ris} functions, such as maintenance, management, and operational tasks, yielding \ac{lifi} networks to be autonomous, prescriptive, and predictive.  Although it is envisioned that the fusion of \ac{ml} and \ac{ris} will evolve the nature of \ac{lifi} applications emerging in all industries,  such as healthcare, retail, transportation, etc., conventional \ac{ml} algorithms are not well suited to guarantee real-time user and network needs in highly dynamic and ultra low-latency-driven  applications \cite{Mohjazi}. Therefore, it is essential to explore innovative mechanisms to resolve the shortcomings of existing \ac{ml} approaches, such as prohibitive training and communication overhead and large processing delays, and hence, open up avenues for exciting applications across all verticals.

\subsection{RIS-Assisted Hybrid \ac{lifi}-RF Networks}
Hybrid \ac{lifi}-\ac{rf} systems have recently emerged as a paradigm-shifting technology that is receiving significant attention,  as they  show outstanding capabilities in enhancing capacity, throughput,  and coverage. This can be attributed to the mutual benefits gained by utilising both the optical and \ac{rf} spectra, such as \ac{rf} traffic offloading, increased \ac{los} blockage mitigation, and improved security and privacy \cite{Ayyash}.  With \ac{ris} being widely recognised as a disruptive approach to augment \ac{rf} communications, the coexistence of \ac{ris}-assisted \ac{lifi} and \ac{ris}-assisted \ac{rf} networks seems to be a natural continuum to hybrid \ac{lifi}-\ac{rf}. This close integration can offer significant synergies, yielding a combined solution that can potentially meet the stringent requirements of future wireless applications, which are defined by ultra-reliable, extremely low-latency, data-driven, and seamless wireless connectivity. Nonetheless, distinctive challenges arise in this case, such as the need to develop complex load balancing mechanisms, reliable resource utilisation  and energy management schemes, efficient cross-layer analytical tools,  and novel cross-band selection combining approaches. One critical question  immediately emerges: what \ac{ml} tools can shake hands with \ac{ris}-assisted hybrid \ac{lifi}-\ac{rf} systems to support real-time dynamic network decision making?

\section{Conclusion}
The merit of \ac{ris} technology opens the door for a whole new realm of wireless applications in which the propagation medium  is  no longer an impediment, but rather an additional degree of freedom.  This article discussed how the interplay between \ac{ris} and \ac{lifi} can lead to innovative and progressive applications by enabling intelligence in the transceiver's operations as well as in the  environment. More research on the deployment of \ac{ris} in \ac{lifi} is fundamental to understand their capabilities and  limitations, and to address the practical considerations that are required to translate this disruptive concept into real-world applications.

\bibliographystyle{IEEEtran}
\bibliography{LiFi}

\balance

\end{document}